\documentclass[12pt,preprint,preprintnumbers,amsfonts,aps,nofootinbib]{article}
\pdfoutput=1

\usepackage{amssymb}
\usepackage{amsmath}
\usepackage{epsfig}

\usepackage{epsf}
\usepackage{graphicx}
\usepackage{amsfonts}
\usepackage{amssymb}
\usepackage{cite}

\def\ie{{\em i.e.\/}}  
\def\eg{{\em e.g.\/}}  
\def\cf{{\em cf.\/}}

\makeatletter
\renewcommand\section{\@startsection {section}{1}{\z@}%
                                 {-3.5ex \@plus -1ex \@minus -.2ex}
                                   {2.3ex \@plus.2ex}%
                                   {\normalfont\large\bfseries}}
\renewcommand\subsection{\@startsection{subsection}{2}{\z@}%
                                   {-3.25ex\@plus -1ex \@minus -.2ex}%
                                     {1.5ex \@plus .2ex}%
                                     {\normalfont\bfseries}}
\renewcommand\subsubsection{\@startsection{subsubsection}{3}{\z@}%
                                   {-3.25ex\@plus -1ex \@minus -.2ex}%
                                     {1.5ex \@plus .2ex}%
                                     {\normalfont\itshape}}
\makeatother

\def\beq{\begin{equation}}
\def\eeq{\end{equation}}
\def\be{\begin{equation}}
\def\ee{\end{equation}}
\def\bea{\begin{eqnarray}}
\def\eea{\end{eqnarray}}

\newcommand{\eq}[1]{equation~(\ref{#1})}

\def\m{\mu}
\def\n{\nu}

\def\P{\Phi}

\def\s{\sigma}

\def\a{\alpha}
\def\b{\beta}
\def\d{\partial}
\def\g{\gamma}

\DeclareRobustCommand{\SkipTocEntry}[4]{}

\begin{document}

\begin{titlepage}

\setcounter{page}{1} \baselineskip=15.5pt \thispagestyle{empty}

\begin{flushright}
\end{flushright}
\vfil

\begin{center}

{\Large \bf   Nonsingular Schwarzschild-de Sitter Black Hole }
\\[0.9cm]
{\normalsize  Damien A. Easson}
\\[0.3 cm]
{\normalsize {\sl  Department of Physics,
Arizona State University, Tempe, AZ 85287-1504}}\\

\vspace{.3cm}

\end{center}

\vspace{.8cm}

\hrule \vspace{0.3cm}
{\small  \noindent \textbf{Abstract} \\[0.3cm]
\noindent We combine notions of a \it maximal \rm curvature scale in nature with that of a \it minimal \rm curvature scale to construct a non-singular Schwarzschild-de Sitter black hole. We present an exact solution within the context of two-dimensional dilaton gravity. For a range of parameters the solution approaches Schwarzschild-de Sitter at large values of the radial coordinate, asymptotically approaching a de Sitter metric with constant minimal curvature, while approaching a maximal constant curvature smooth spacetime as the radial coordinate approaches zero. The spacetime is geodesically complete and generically has both a black hole horizon and a cosmological horizon. }\vspace{0.5cm}  \hrule

\vfil

\begin{flushleft}
{\normalsize { \sl \rm \small{easson@asu.edu}}}\\
\end{flushleft}

\end{titlepage}

\newpage
\tableofcontents

\section{Introduction}
There are few puzzles in theoretical physics more vexing than the enigmatic singularity on the interior of a black hole. This physical impossibility, is assumed to be conveniently hidden from the distant observer by an event horizon. According to Einstein's theory of General Relativity (GR), an observer falling into the black hole will inevitably encounter ever-increasing destructive tidal forces on his way towards the singularity, a fate guaranteed by the singularity theorems of Hawking and Penrose~\cite{Hawking:1969sw}. The singularity event occurs at extremely high curvatures, near the Planck scale. At such a high scale it is known that General Relativity breaks down, leading many physicists to speculate that the singularity might somehow be tamed by higher order corrections to GR or quantum gravitational effects. 

One, now well-known, idea of how the singularity might be resolved is realized in the \emph{ Limiting Curvature Hypothesis} (LCH)\cite{markov}. Markov's hypothesis rests on the fact that, given the existence of a fundamental natural length scale $\ell_f$, all curvature invariants will be bounded, \eg~the Kretschmann scalar $R_{\a\b\g\delta}R^{\a\b\g\delta} \leq \ell_f^{-4}$. Bounding all curvature invariants ensures the absence of curvature singularities in the theory. In addition, if the corresponding limiting curvature is appreciably below the Planck scale, any quantum effects can be safely ignored.  As a result, the LCH has been used to remove not only black hole singularities but cosmological big-bang curvature singularities as well
\cite{Frolov:1989pf,Frolov:1988vj,Morgan:1990yy,Mukhanov:1991zn,Trodden:1993dm,Brandenberger:1993ef,Moessner:1994jm,Brandenberger:1995es,Brandenberger:1998zs,Easson:1999xw,Easson:2002tg,Easson:2003ia,Chamseddine:2016uef,Chamseddine:2016ktu}.

In this paper we follow the LCH construction of a nonsingular black hole in 2D dilaton gravity found in \cite{Trodden:1993dm}. However, in addition to a maximal limiting curvature we apply the minimimal curvature conjecture (MCC) introduced in \cite{Easson:2006jd}. By limiting the curvature at both extrema we find an exact solution corresponding  to a nonsingular Schwarzschild-de Sitter (SdS) black hole. 
Two-dimensional dilaton gravity is the most general action for gravity with a scalar field in two-dimensions (see \eg~\cite{Banks:1992xs}).  It is only natural to question the physical validity of a 2D theory; however, there are strong motivations for using the theory to capture possible aspects of quantum gravity, and significant progress has been made in 2D theories that recover aspects of higher-dimensional gravitational theories. Furthermore, our two dimensional theory allows for a completely analytic treatment and it is our hope the solution captures non-perturbative features of the physics, evaporation, and the final state of realistic four-dimensional black holes. This hope is bolstered by past work. For example, It has been established that many thermodynamic properties of black holes in higher dimensions can be captured by the thermodynamics of black holes in 2D dilaton gravity \cite{Banks:1992xs,Youm:1999xn}. Specifically, it has been shown that the Bekenstein-Hawking (BH) entropies of dilaton black holes and p-branes are exactly reproduced by the BH entropy of 2D dilaton black holes. Further motivation is provided by the successful generalization of a nonsingular cosmology in two-dimensions \cite{Moessner:1994jm}, to a nonsingular cosmology in four-dimensions \cite{Brandenberger:1998zs}.  Inspired by these successes, it is our hope this work will serve as a stepping stone towards the construction of a more realistic 4D solution.

Altering the traditional black hole solution by removing the singularity has important consequences for the early universe, the nature of dark matter, Hawking radiation and associated phenomena such as firewalls and the information loss problem. We return to these considerations below.

\section{Two Dimensional Gravity}
Our starting point is a $D=1+1$ dimensional, dilaton gravitational theory. This simple yet rich theory allows for exact analytic computation of our solution. In addition to its tractability, several aspects of four-dimensional theories can be captured in the two-dimensional case, including aspects of perturbative quantum gravity and black hole thermodynamics (for a review see~\cite{Grumiller:2002nm}). The most general Lagrangian in $2D$ for gravity and a scalar is \cite{Banks:1992xs}:
\be\label{srgrav}
S = \int d^{2} \! x \,  \sqrt{-g} \, \Big( \mathcal{D}( \Phi) R   \\ \nonumber
 +   \mathcal{G}(\Phi) \, (\nabla \Phi)^2
                   + \mathcal{H}(\Phi) \Big )
\,,
\ee
were $g$ is the determinant of the metric tensor $g_{\m\n}$, $R$ is the Ricci scalar and $\Phi$ is the dilaton scalar field. 
After the Weyl rescaling 
\be
g_{\m\n} \rightarrow e^{2 \s(\Phi)} g_{\m\n}
\,,
\ee
and requiring that 
\be
4\frac{d\s}{d\Phi} \frac{d \mathcal{D}}{d\Phi} = - \mathcal{G}(\Phi)
\,,
\ee
it is possible to conformally transform away the kinetic term for $\Phi$:
\be\label{start}
S=\int d^2x \sqrt{-g} \left(\mathcal{D}(\Phi) \,R + \mathcal{V}(\Phi) \right)
\,,
\ee
where $\mathcal{V}(\Phi)=e^{2 \s(\Phi)} \mathcal{H}(\Phi)$.

Varying this action with respect to $\Phi$ and the metric tensor $g_{\m\n}$ yields
\bea
- \frac{\d \mathcal{V}(\Phi)}{\d \Phi} &=& \frac{\d \mathcal{D}(\Phi)}{\d \Phi} R \label{eo1} \\
\mathcal{V}(\Phi)\,g_{\m\n} &=& 2(g_{\m\n} \nabla^2  - \nabla_\m \nabla_\n) \mathcal{D}(\Phi)
\,,
\eea
respectively.  We now choose $\mathcal{D}(\Phi)=1/\Phi$.

\section{Nonsingular SdS Black Hole Solution}

We assume a
static metric ansatz with timelike Killing vector $K^\m = (\d_t)^\m$
\be\label{ds2d}
ds^2 = -n(r)dt^2 + p(r)dr^2
\,.
\ee
The equations of motion may be obtained by varying the action with respect to $n$, $\Phi$ and $p$ respectively:

\bea\label{eomset}
p^2 \Phi^3 \mathcal{V}(\Phi) - 4 p \Phi'^2 - p'\Phi'\Phi + 2 p \Phi \Phi'' &=& 0 \\
p n'^2 - n n' p' + 2 n^2 p^2 \Phi^2 \frac{\d \mathcal{V} }{\d \Phi} - 2 n p n'' &=& 0 \\
n p \Phi^2 \mathcal{V}(\Phi) +\Phi' n' &=& 0
\,.
\eea

Choosing the ``Schwarzschild gauge" $p(r)=1/n(r)$, the EOM become
\bea\label{eom}
\Phi^3 \mathcal{V}(\Phi) - 4 n \Phi'^2 + n'\Phi'\Phi+ 2 n \Phi \Phi'' &=& 0 \\
\frac{\d \mathcal{V} }{\d \Phi} + \Phi^{-2} n'' &=& 0 \\
\mathcal{V}(\Phi) + \Phi^{-2} \Phi' n' &=& 0 \label{eom33}
\,.
\eea
\subsection{Extrema Curvature Conjecture}
In accordance with the limiting curvature hypothesis we now identify a class of potentials 
$\mathcal{V}(\Phi)$ that will ensure $R$ remains bounded for all values of the radial coordinate $r$ from
$0$ to $\infty$.  As $R$ is the only curvature invariant in $D=2$, the LCH now ensures the spacetime is free of curvature singularities. 

The form of the potential is restricted by considering the behavior of the system at low and high curvatures. 
We now assume the existence of a minimal curvature scale $R=R_{min}$, as in the minimal curvature conjecture \cite{Easson:2006jd}.
In a four-dimensional cosmology with flat Friedmann-Robertson-Walker (FRW) universe
with metric 
\be
ds^2 = -dt^2 + a^2(t) dx^i dx_i
\,,
\ee
the minimal curvature scale may be parameterized by a cosmological constant $\Lambda$, and is given in terms of the Hubble parameter $H = \dot a/a$, by
$R_{min} = 12 H^2 = 4 \Lambda$.
In the SdS spacetime under consideration, we asymptotically approach a minimal curvature de Sitter spacetime at infinity, and the desired solution for $n(r)$ should reduce to the Schwarzschild de-Sitter solution,  
requiring
\be\label{ninf}
n(r)\rightarrow 1-\frac{2m}{r} - \frac{\Lambda r^2}{3} \qquad \mathrm{when} \qquad r\rightarrow \infty, \; \Phi \rightarrow 0
\,.
\ee
At this point it is instructive to consider the structure of an ordinary, singular, Schwarzschild-de Sitter spacetime. 
For this purpose, a Penrose diagram for the SdS solution is presented in Fig.~1.
\begin{figure}\label{penrose.png}
\centerline{\includegraphics[width=4.75in]{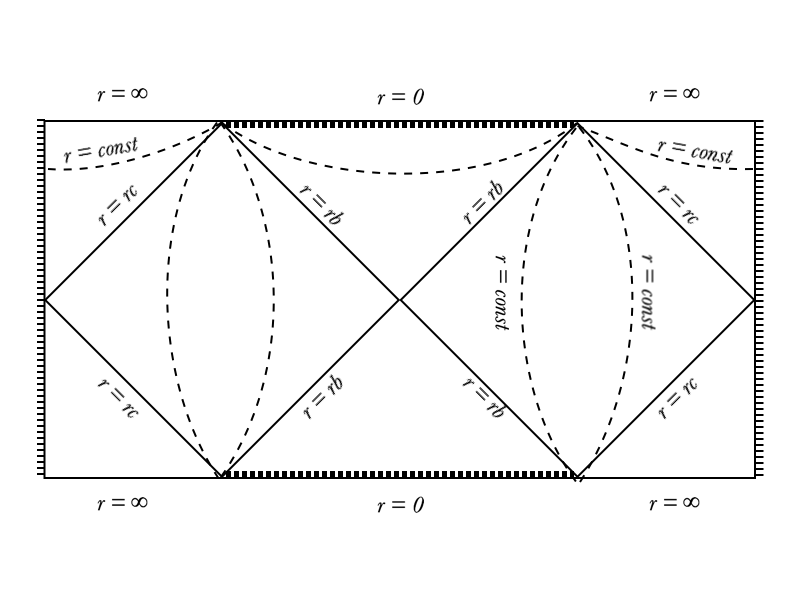}}
\caption{Conformal diagram for Schwarzschild-de Sitter spacetime. There are analytic continuations to the left and right of the diagram. Thus, to avoid an infinity of black hole and white hole singularities (horizontal dashed lines at $r=0$) one typically identifies
along the left and right vertical dashed lines to achieve a single white hole singularity in the past and a single black hole singularity in the future. The spacetime has both a black hole horizon at $r=r_b$ and a cosmological horizon located at $r=r_c$. In the model under consideration the singularity at $r=0$ will be replaced by a smooth spacetime of constant maximal curvature.}
\end{figure}

In addition to the minimal curvature we make a Markovian assumption of a maximal, high energy curvature scale $R_{max}$, such that $R\leq R_{max}$ \cite{markov}, which is reached
deep inside of the black hole, replacing the singularity at $r=0$. 
Using our aforementioned choice for $\mathcal D$,  \eq{eo1} becomes $R = \Phi^2 \d \mathcal{V}(\Phi)/\d \Phi$  and it is easy to see that
limiting $\partial \mathcal{V}/\partial \Phi$ in the high curvature regime (\ie \, $r \rightarrow 0$, $\Phi \rightarrow \infty$)
will result in the limitation of $R$. 

\subsection{Interpolating potential construction}
With the anzats $\Phi = 1/\a r$, and using the above relation for $R$, we find the low-curvature asymptotic behavior of 
the potential
\be\label{vlarge}
\mathcal V(\Phi) \simeq 2 m \a^3 \Phi^2 - 2 \Lambda/3 \Phi \qquad \mathrm{when} \qquad r\rightarrow \infty, \; \Phi \rightarrow 0 \,.
\ee
In the high curvature regime $R \rightarrow R_{max}$, we choose to parametrize the maximum curvature by $\ell$, so that $R_{max} \sim 2/\ell^2$. In this regime, integrating \eq{eom33}, leads to:
\be
n(r)\rightarrow \frac{r^2}{\ell^2} - const \qquad \mathrm{when} \qquad r\rightarrow 0, \; \Phi \rightarrow \infty
\,.
\ee
Since $\Phi \rightarrow \infty$,  we see that
$\d \mathcal{V}(\Phi)/\d \Phi \propto \Phi^{-2}$, and therefore 
\be\label{vsmall}
\mathcal{V}(\Phi) \simeq \frac{2}{\ell^2} \frac{1}{\Phi}   \qquad \mathrm{when} \qquad r\rightarrow 0, \; \Phi \rightarrow \infty \,.
\ee
The following potential correctly interpolates between the above asymptotic regimes (\ref{vlarge}) and (\ref{vsmall}),
\be
\mathcal{V}(\Phi) = \frac{2 (m \a^3 \Phi^2 - \Lambda/3\P)}{1 + m \a^3 \ell^2 \Phi^3}
\,,
\ee
While our particular $\mathcal D$ and $\mathcal{V}$ are not unique, they satisfy the appropriate asymptotic behaviors and lead to the unique exact solution (\ref{nofr}). 

\subsection{Solving the equations of motion}
Using this potential in the EOM, \eq{eom33} becomes
\be
n'(r) = \frac{2( 3 m r - r^4 \Lambda)}{3(\ell^2 m + r^3)} \,,
\ee
which may be integrated to find the exact analytic solution

\bea
\Phi(r) &=& \frac{1}{\a r}  \,, \qquad \qquad \qquad \qquad \qquad \;\;\;\;\;\;\\
n(r)
& = &   \frac{1}{3} \Big(\frac{m}{\ell}\Big)^{2/3} \Big(1+   \frac{ \ell^2 \Lambda}{3}  \Big)\,
\ln {\Big{[}\, \frac{r^2 - (m \ell^2)^{1/3}r + \nonumber
(m\ell^2)^{2/3}}{r_0^2 - (m\ell^2)^{1/3}r_0 + (m \ell^2)^{2/3}} \Big(\frac{r_0 + (m\ell^2)^{1/3}}{r + (m\ell^2)^{1/3}}\Big)^2 \, \Big]} \nonumber \\
	&+ &  \frac{2}{\sqrt{3}} \Big(\frac{m}{\ell}\Big)^{2/3}\,\Big(1+ \frac{ \ell^2 \Lambda}{3} \Big)
	\Big\{\operatorname{arctan}\Big(\frac{2r - (m\ell^2)^{1/3}}{\sqrt{3} (m\ell^2)^{1/3}}\Big) \nonumber
	-\operatorname{arctan}\Big(\frac{2r_0 - (m\ell^2)^{1/3}}{\sqrt{3} (m\ell^2)^{1/3}}\Big)\Big\} \\
	&-&  \frac{\Lambda}{3}  (r^2 - r_0^2)  \label{nofr}
\,,
\eea
where we have introduced $r_0$ for the location of the black hole horizon. The function $n$ is plotted in Fig.~2, for various parameter values.  In general the solution has a black hole horizon and a cosmological de Sitter horizon. Outside of the black hole horizon, for large $r$, the solution asymptotically approaches de Sitter space at the minimal limiting curvature scale given by $\Lambda$. Inside the black hole horizon the solution remains everywhere nonsingular and approaches a spacetime of constant maximum limiting curvature at $r=0$ parameterized by $\ell$. 
\begin{figure}\label{nofr.png}
\centerline{\includegraphics[width=4.75in]{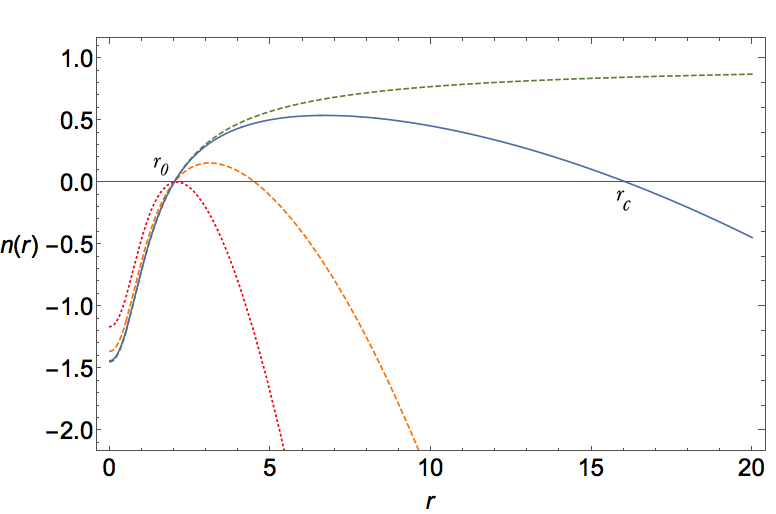}}
\caption{Plot of  (\ref{nofr}), the metric function $g_{00} = n$ as a function of the radial coordinate $r$. In the plots we set $m$ and $\ell$ to unity and allow $\Lambda$ to vary. The spacetime has a black hole horizon at $r=r_0$ and a cosmological horizon located at $r=r_c$ (solid, blue solution, $\Lambda = .01$). By adjusting the value of $\Lambda$, the value of $r_c$ can be made arbitrarily large. There is a critical ratio for $\Lambda$ and $m$ where the two horizons meet (lower dotted, red solution, $\Lambda=.3$) corresponding to the extremal Schwarzschild-de Sitter solution. Setting $\Lambda =0$, reduces our large $r$ solution for $n$ (\cf~\ref{ninf}) to the Schwarzschild solution and hence approaches unity at large $r$ (upper dashed, green solution). The arbitrary solution in dashed orange has $\Lambda = .1.$}
\end{figure}
This interior solution is conveniently described by the coordinates in Appendix~\ref{coord}. The successful implementation of the extrema curvature conjecture is clearly visualized in a plot of the Ricci scalar $R$ for the entire range of the radial coordinate (see Fig.~3). Note the interesting feature that curvature scalar changes sign. This sign change occurs inside the event horizon and therefore is hidden from the external observer. Thus the entire solution interpolates between spacetimes of different constant curvatures  at the center and at infinity, separated by the black hole horizon. Further implications of this (and the evaporation process) will be given in \cite{Easson:2001qf}.

For the curious reader, the Ricci scalar computed from the metric (\ref{ds2d}) is:
\be
R = \frac{n n' p' + p (n'^2 - 2 n n'')}{2 n^2 p^2}
\,.
\ee
With the choice $p=1/n$, together with the solution for $n$ given by (\ref{nofr}) the curvature becomes:
\be\label{riccirp}
R = - n'' =  \frac{2(6 m r^3 + \Lambda r^6 + \ell^2 m (4 r^3 \Lambda - 3 m))}{3 (\ell^2 m + r^3)^2} \,.
\ee

We emphasize, for the reader who may consider the curvature conjecture to be speculative, that the black hole solution above is an exact solution to the dilaton gravity theory. The conjecture served merely as a guiding principle to arrive at the desired solution.

\begin{figure}
\centerline{\includegraphics[width=4.75in]{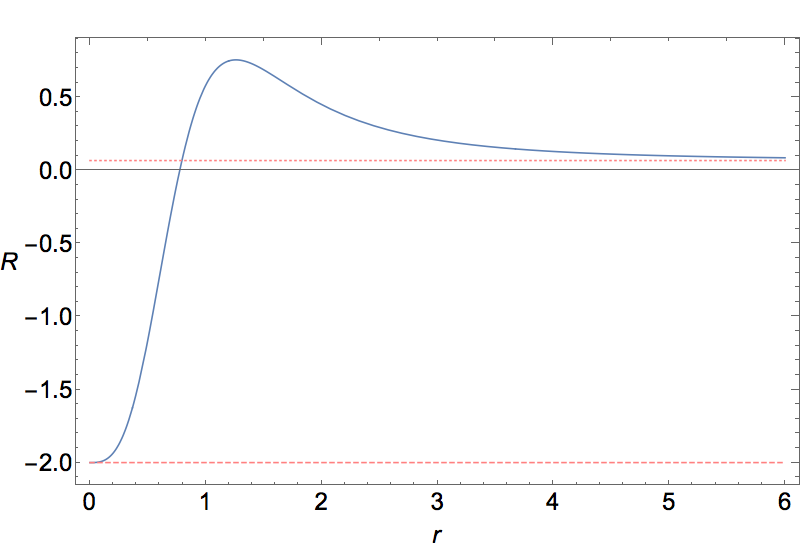}}
\caption{Plot of scalar curvature $R$ (\ref{riccirp}) as a function of the radial coordinate $r$. Parameter values shown are $\Lambda = 0.1$, $m = 1$, $\ell = 1$. For large values of the 
radial coordinate the minimal curvature hypothesis ensures the curvature approaches the minimal curvature de Sitter spacetime with Ricci scalar $R_{min} = 2 \Lambda/3$ (upper dotted line). As $r \rightarrow 0$, the would-be
black hole singularity is replaced by a smooth spacetime with constant maximal curvature  $R_{max} = -2/\ell^2$ (lower dashed line).}
\end{figure}

\section{Conclusions}
In this paper we have suggested an \emph{extremal} curvature conjecture--combining the traditional maximal curvature conjecture \cite{markov} with the minimal curvature conjecture of \cite{Easson:2006jd}. Applying this principle to bound curvatures at both UV and IR scales, we find a new static,  geodesically complete spacetime solution in two-dimensional dilaton gravity. The exact analytic solution corresponds to a nonsingular, double-horizon Schwarzschild-de Sitter black hole. Removing the black hole singularity has drastic consequences for the final nature of black hole evaporation as shown in \cite{Easson:2002tg}. There, the evaporation of the nonsingular black hole constructed in \cite{Trodden:1993dm} was analyzed. Removal of the Schwarzschild singularity resulted in an eternally slowly radiating black hole remnant, suitable as a dark matter candidate. In addition, the black hole remnant has ample room on the interior to store information, leaving a possible resolution to the black hole information loss paradox.~\footnote{A complimentary analysis using the limiting maximal curvature hypothesis applied to a four dimensional nonsingular black hole was recently conducted in  \cite{Chamseddine:2016ktu}.}
In the present case the nonsingular Schwarzschild-de Sitter black hole has two horizons, and we leave a complete understanding of the evaporation process for future study~\cite{pdde}. The model considered here corroborates the ubiquitous notion found in the literature that a universe may be born on the interior of a black hole. A realistic model of this sort, perhaps along the lines of the discussion in \cite{Easson:2001qf}, remains an open challenge.


\appendix
\section{Coordinates}\label{coord}
Near $r=0$ and the would-be black hole singularity, we may expand $n(r)$ (\ref{nofr}) to second order:
\be
n(r) \simeq  \frac{r^2}{\ell^2} - const + \mathcal{O}(r^3) \,,
\ee
and setting the constant to unity the metric as $r \rightarrow 0$ becomes
\be\label{oldle}
ds^2 = - \left( \frac{r^2}{\ell^2} - 1 \right) dt^2 + \frac{dr^2}{ \left( \frac{r^2}{\ell^2} - 1 \right) } \,.
\ee
Introducing the new coordinate
\be
\chi = \int \left(\frac{r^2}{\ell^2} - 1 \right) ^{-\frac{1}{2}} \, dr \,,
\ee
so that after appropriate rescaling
\be
\frac{\chi}{\ell} = \ln{\Bigg[ \frac{r}{\ell}  + \sqrt{  \frac{r^2}{\ell^2} - 1}} \,\Bigg] \, .
\ee
In terms if the new coordiate $\chi$, the line element (\ref{oldle}) becomes
\be
ds^2 = - \operatorname{sinh}^2 \left(\frac{\chi}{\ell}\right)  dt^2 + d\chi^2
\,.
\ee


\begingroup\raggedright


\end{document}